\documentclass[aps, prd,showkeys]{revtex4}

\usepackage{amsfonts}       
\usepackage{graphicx}
\usepackage{amsmath}
\usepackage{amssymb}

\usepackage{color}

\newcommand{\bld}[1]{\mathrm{\textbf{#1}}}
\newcommand{\Tr}[1]{\mathrm{Tr}\, \mathrm{#1}}
\newcommand{\mat}{\mathbb}
\newcommand{\beq}{\begin{equation}}
\newcommand{\eeq}{\end{equation}}
\newcommand{\barr}{\begin{eqnarray}}
\newcommand{\earr}{\end{eqnarray}}

\newcommand{\Un}[1]{\, \mathrm{#1}}
\newcommand{\lf}{\left\lfloor}
\newcommand{\rf}{\right\rfloor}
\usepackage{soul}
\usepackage[normalem]{ulem}
\begin{document}

\title{The Axial Anomaly in Lorentz Violating Theories: Towards the Electromagnetic Response of Weakly Tilted Weyl Semimetals}

\author{Andr\'es G\'omez$^a$}
\email{andresgz@ciencias.unam.mx}
\affiliation{$^a$Facultad de Ciencias, Universidad Nacional Aut\'onoma de M\'exico, \\
Circuito Exterior, C.U., 04510 CDMX, M\'exico}

\author{Luis  Urrutia$^b$}
\email{urrutia@nucleares.unam.mx}
\affiliation{$^b$Instituto de Ciencias Nucleares, Universidad Nacional Aut\'onoma de M\'exico, \\
Circuito Exterior, C.U., 04510 CDMX, M\'exico}
\begin{abstract}
Using the path integral formulation in Euclidean space, we extended the calculation of the abelian chiral anomalies in the case of Lorentz violating theories  by considering a new fermionic correction  term provided by the standard model extension, which arises in the continuous Hamiltonian of a weakly tilted Weyl semimetal, and whose cones have opposite tilting. We found that this anomaly is insensitive to the tilting  parameter, retaining its well-known covariant form.
This independence on the Lorentz violating parameters is consistent with other findings reported in the literature.
The initially imposed gauge invariant regularization was consistently recovered at the end of the calculation by the appearance of highly non-trivial combinations  of the covariant derivatives, which ultimately managed to give only terms  containing the electromagnetic tensor. 
We emphasize that the value of the anomaly with an arbitrary parameter is not automatically  related  to the effective action describing the electromagnetic response of such materials.   
\end{abstract}

\maketitle

\section{Introduction}
\label{INTRO}

 The  appearance of fermionic excitations  in the  continuum Hamiltonians of condensed matter materials, which naturally violate some spacetime symmetries  and incorporate contributions similar to those in the fermionic sector of the standard model extension (SME) \cite{SM1,SM2}, has created the possibility of applying  many  of the techniques  already developed in the search for a fundamental Lorentz violation in the interactions of high energy physics to this area. This also occurs  in the effective electromagnetic response of magnetoelectic materials, for~example,  which  also reproduce some  of the terms included in the  electromagnetic sector of the  SME~\cite{SM3,SM4}.
 
Macroscopic electromagnetic transport properties in condensed matter are often calculated using the Kubo formula~\cite{KUBO} or the semiclassical Boltzmann approach~\cite{Boltzmann}. Recently, the~rise of topological phases of matter has promoted  the use of anomaly calculations for these purposes. The~topological properties of anomalies have long been recognized in high energy physics and their presence is expected to provide a macroscopic understanding of the underlying topological properties in the electronic design of these materials. Such an approach has been particularly fruitful in the case of Weyl semimetals (WSMs)  whose electronic  Hamiltonians naturally include some of the Lorentz invariance violating (LIV) terms considered  in the fermionic sector of the standard model extension (SME) \cite{SM5}.
 Nevertheless, in~this case, the LIV parameters need not be highly suppressed, since they are determined by the electronic structure of the material and subjected to experimental determination.   
 Weyl semimetals were  first theoretically predicted
in pyrochlore iridates (such as ${\rm Y}_2{\rm Ir}_2{\rm O}
_7$) in 2011~\cite{XWAN} and
experimentally discovered in TaAs four years later~\cite{SMHUANG,BQLV,SYHU,LXYANG,SYHU1}.  
Their  low energy excitations  near the Fermi energy are described by Weyl fermions whose band structure is characterized by an even number of gapless Weyl  nodes with opposite chiralities, which are the only points where the valence band touches the conduction band. This pair-wise characteristic is a consequence of the Nielsen--Ninomiya theorem~\cite{NN} and the stability of the Weyl cones is guaranteed by the breaking of time reversal and/or inversion symmetries.  Each pair of nodes can present a separation in
energy and/or momentum in the Brillouin zone.  Fixing one of the momenta in the Brillouin zone, say $k_z$, the~dispersion relation of the Weyl fermions is  $E(\mathbf{k})=\pm \sqrt{k^2_x + k^ 2_y} \,$ for the energy spectrum near the nodes. This surface describes  a cone with an apex at the origin and axis perpendicular to the  $(k_x-k_y)$ plane where we fix the Fermi energy touching the cone just in the~apex. 

A simple model of a WSM consisting of two Weyl nodes is described by the fermionic action:
\beq
S=\int d^4 x \Big( i\bar{\Psi}\left( \gamma ^{\mu }\partial _{\mu}+ib_{\mu }\gamma ^{\mu
}\gamma ^{5}\right) \Psi \Big),
\label{actionWSM}
\eeq 
where $b_\mu$ signals a Lorentz violating term which accounts for the separation of the Weyl nodes in energy ($b_0$) and in momentum (${\mathbf b}$). Adding an additional Lorentz violating term  in the derivative term of the action (\ref{actionWSM}) makes it possible to give an inclination to the cone axis, thus defining a tilted WSM. For~small inclinations (small tilting), we keep the  point-like Fermi surface and these materials are called Type-I WSMs. By increasing the tilting, we arrive at a point where the conical surface becomes tangent to the $(k_x-k_y)$ plane. A~further increment  will produce an intersection of the conical surface with the Fermi energy plane creating what are called the electron and hole pockets, which takes us out from the point-like Fermi surface. This second possibility gives rise to Type II  WSMs~\cite{Trescher,Soyulanov}.

Among the novel transport properties of WSMs, we find the chiral magnetic effect, whereby a ground state dissipationless current proportional to an applied magnetic field is generated in the bulk of a WSM with broken inversion symmetry~\cite{FUKUSHIMA}.  This yields a conductivity proportional to the magnetic field, or~equivalently, to a resistivity that decreases with an increasing magnetic field. This phenomenon, dubbed as negative magnetoresistance,  was predicted in Ref.~\cite{NN} and was experimentally observed in Ref.~\cite{KIM}. WSMs also exhibit the anomalous Hall effect,  characterized by Hall conductivity proportional to the separation of the Weyl nodes in momentum~\cite{BB, GRUSHIN, ZYUZIN, GOSWAMI}.  For~a review of WSMs, see, for example, Refs.~\cite{REV1,REV2,REV3}.

The appearance of  Weyl fermions quasi-particles in WSMs naturally introduces the issue of anomalies when gauge invariance is demanded via the minimal coupling with external electromagnetic fields which is  required to probe the electromagnetic response of such materials~\cite{VAZIFEH}. In~this way, the chiral current $J_\mu^5={\bar \Psi} \gamma^\mu \gamma^5 \Psi$ is not conserved, yielding the abelian version of the chiral anomaly:

\beq
\partial^\mu J_\mu^5= -\frac{e^2}{16 \pi ^2}\epsilon^{\mu\nu \alpha \beta}F_{\mu\nu}F_{\alpha\beta} = +\frac{e^2}{2 \pi^2} {\mathbf E}\cdot {\mathbf B},
\label{INTRO_AA}
\eeq

in the Lorentz covariant case \cite{ADLER,BJ}.  Here we follow the conventions of Ref. \cite{JACKSON} with $\epsilon^{0123}=\epsilon^{1234}=+1$ .
In the path integral approach, the~effective macroscopic electromagnetic action is obtained by introducing the electromagnetic coupling in  \mbox{(\ref{actionWSM})} and subsequently 
trading the fermionic term ${\bar \Psi}b_\mu \gamma^\mu \gamma_5 \Psi $
by an electromagnetic contribution arising from 
the Jacobian of the chiral transformation, which eliminates the $b_\mu$ term from the action~\cite{ZYUZIN}.
 Following this idea, the~Fujikawa prescription to calculate the chiral anomalies~\cite{FUJIKAWA,BERTLEMENT} has  been extensively  used for  these purposes~\cite{ZYUZIN,GOSWAMI}.

 The relevance of the tilting of the cones in the transport properties of WSMs was first reported in Ref.~\cite{Trescher}. Subsequent works aiming to understand the role  of the chiral anomaly in this process were based on the semiclassical Boltzman approach. Here, the contribution of the anomaly was identified through the term $\mathbf{E}\cdot \mathbf{B} $  which appears as a factor of the Berry phase in the equation of motion of the momentum of the wave packet, which is finally solved  in terms of the external electromagnetic fields~\cite{SHARMA, ZYUZIN1}.

 In this work, we take the first steps in calculating the effective electromagnetic action  of an isotropic tilted  Type I WSM, by considering a model with a single pair of Weyl points with tilting in opposite directions.  Using  the path integral approach, we deal with the calculation of the anomaly corresponding to the modified axial current responsible for the tilting of the cones around each node, which amounts to discarding  the effect of the $b_\mu$ contribution in a first approximation. In~this restricted setting, our work is similar to the calculation of the anomaly presented in Ref.~\cite{KZ}, however,~we find some important differences arising from the non-commutativity of the operators involved in the calculation, which we report in Section~\ref{DISC}. We only consider the corrections to  the anomaly which are linear and quadratic in the tilting~parameter.  

 This paper is organized as follows. In~Section~\ref{MODDL}, we define  our model and provide a brief explanation of WSMs, setting  the notation and conventions. Section~\ref{AXIALAN} deals with a review of the axial anomaly in the path integral approach when going to  Euclidean space. The~modified axial current is calculated and the basic operators required for the 
 implementation of the Fujikawa method are derived. We then proceed to the first  order  calculation (linear in the tilting parameter) in Section~\ref{First_Order}, followed by the second order  calculation  (quadratic in the tilting parameter) in Section~\ref{Second_Order}. We conclude with a summary and a discussion of the results in Section~\ref{DISC}. In~this section,
 we also make a comparison with previous findings in the literature.  The~Appendix  \ref{USEREL}  summarizes some useful relations used in the~calculations.

\section{The Modified Dirac~Lagrangian}
\label{MODDL}
In this article, we consider the Lagrangian:
\begin{equation}
\label{Lagrangian}
    \mathcal{L} = \bar{\Psi}(x)i \mat{D}\Psi(x) = i\bar{\Psi}(x)\big(\gamma^{\mu}D_{\mu}+\gamma^0\gamma^5v^iD_i\big)\Psi(x),
\end{equation}

which describes the dynamics of a massless spinor field $\Psi(x)$, coupled with an external electromagnetic field $A_\mu$ through the covariant derivative $D_{\mu} = \partial_{\mu}-ieA_{\mu}$. The~standard gamma matrices $\gamma^\mu$  would ensure the Lorentz invariance of the Lagrangian if it were not for the presence of the second term. This term contains the pseudoscalar matrix $\gamma^5 = i\gamma^0\gamma^1\gamma^2\gamma^3$ and a Lorentz violating parameter $v^i, \, i=1,2,3$,  which  we consider  as the components of  a spacetime vector $\bld{v}$.  In~this way, we keep invariance under rotations. For~$\bld{v}=0$, we recover the usual Dirac Lagrangian for massless~spinors. {As is usual, $\bar{\Psi}(x)=\Psi^{\dagger} \gamma^0.$}

The physical motivations for taking the modified Lagrangian (\ref{Lagrangian}) can be seen from two branches of physics. On~one hand, in~the description of  Weyl semimetals, $\bld{v}$ parametrizes a tilt of the cones in a system with a pair of  Weyl nodes. This parameter, which is  determined by the electronic structure of the material,  could in principle  take any value, as~Lorentz invariance is effectively  broken into condensed matter. On~the other hand, in~the context of quantum field theory, this term probes the possibility of a fundamental breaking of Lorentz invariance. As~this possibility has not been found in the many high-precision experiments already performed, we must consider this parameter as very small in this setting.
Indeed, the~Lagrangian (\ref{Lagrangian}) is a particular case of:
\begin{equation}
    \mathcal{L} = \bar{\Psi}(x)\big(i\Gamma^{\mu}\partial_{\mu}-M\big)\Psi(x),
\end{equation}

which arises in the study of possible Lorentz violations in the fermionic sector of the  SME \cite{SM5}, with:
\beq
\Gamma^\mu=\gamma^\mu+ c^\mu{}_\nu \gamma^\nu+ d^\mu{}_\nu \gamma^\nu \gamma^5, \qquad M = m +  b_\mu \gamma^\mu \gamma^5 +\frac{H_{\mu\nu}}{2} \sigma^{\mu\nu}.
\label{GAMMAMU}
\eeq

In particular, we will calculate the abelian axial anomaly for the choice
{$c^\mu{}_\nu=0$ and } $M=0$. We rewrite the Lagrangian (\ref{Lagrangian}) in a way that makes explicit the transformation properties of the indices involved. To~this end, we follow the notation of the Dirac equation in curved space and  write (\ref{Lagrangian}) as 
\beq
\mathcal{L} = \bar{\Psi}(x)i\gamma^{A} e_A{}^\mu\partial_{\mu}\Psi(x),
\eeq
where Latin indices $A=0,1,2,3 : \, \{0, \,  a=1,2,3 \}$ live in the matrix space, while Greek indices $\mu= 
0,1,2,3 : \, \{0, \, i=1,2,3 \}$ label spacetime coordinates. Then, the indices $A$ transform under  local Lorentz transformations while the indices $\mu$ transform under general coordinate transformations. In~our flat space setting, both types of transformations coincide, but~this distinction will be relevant when we make the continuation to Euclidean~space.

 We further constrain ourselves to the case:
\beq
e_A{}^\mu= \delta^\mu_A+ d^\mu{}_A \gamma^5,
\eeq 
and we interpret $d^\mu{}_A$ as four spacetime vectors labeled by the Dirac indices $A$. Moreover, the Lagrangian in  Eq. (\ref{Lagrangian}) yields the further restriction:
\beq
d^\mu{}_A=\delta^\mu_i \delta^0_A v^i,
\eeq
which says that we have only one spacetime vector $v_0^\mu$ corresponding to $A=0$ with only spatial components $v_0^i=v^i$.  The~main point here is that both types of indices have definite transformation properties when changing coordinate frames. In~this notation, the gamma matrices satisfy $\{ \gamma^A, \gamma^B \}= 2 \eta^{AB}$ with $\eta^{AB}=diag(+1,-1, -1, -1)$. The~spacetime metric is $\eta^{\mu \nu}= diag(+1,-1, -1, -1)$.


\subsection{A Brief on Weyl~Semimetals}
\label{Weyl}

In a low energy approximation near the Fermi energy,  Weyl semimetals are materials for which  their band structure can be described by a pair (or an even number) of nodes (Weyl points)  separated in energy and momentum. The~low energy theory of an isolated Weyl point is given by the Hamiltonian~\cite{Soyulanov}:
\begin{equation}
\label{H_weyl}
    \Un{H}_{\pm}(\bld{k}) = \pm \sigma^0 \bld{v}_0 \cdot \bld{k} + \chi\sigma^{a}\bld{v}_a\cdot\bld{k} = \sigma^A v^i_A(\pm, \chi) k_i, \qquad \quad  v^i_A(\pm, \chi)=(\pm v_0^i, \chi  v_a^i),
\end{equation}
where $\sigma^{a}$ ($a=1,2,3$) are the Pauli matrices, $\sigma^0=I$ is the unit $2\times 2$ matrix ,  $\chi = \pm 1$ is the chirality of the node and $ \pm {\bld{v}_0}$ indicates the direction of the tilting. The~Nielsen--Ninomiya theorem~\cite{NN} states that each pair of cones must carry  opposite chiralities.
\mbox{In Eq.(\ref{H_weyl}),}  $\bld{k}$ is the crystal momentum vector centered in each node. As~shown in Ref.~\cite{ZYUZIN} for the untilted case, taking into account the separation of the  nodes has direct consequences on the coefficient of the~anomaly. 

We have four spacetime vectors $\bld{v}_{A} = (v^{i}_A)$ ($A=0,1,2,3$, $i=1,2,3$)  with only spatial components $v_A^i$. That is, for~each particular label $i$, the~object $v_{A}^i$ transforms as a covariant vector in Lorentz space.  The~vectors
$v_{A}^i$ characterize the  cone structure arising from the dispersion relations, where  the apexes of the cones are located at the nodes. The~vectors $\bld{v}_a$ describe  the anisotropy of the cones   while  $\bld{v}_0$ gives their tilting.  Our choice  $\pm\bld{v}_0$ in the Hamiltonian (\ref{H_weyl}) indicates that we are considering the two cones with tilting in the opposite~directions. 

The case of tilting in the same direction has been considered  in Ref.~\cite{KZ} and results in no additional terms in the axial anomaly, yielding a Lorentz invariant result.  
Although the authors in Ref.~\cite{KZ} obtained  a non-trivial anomaly in the case of cones with  opposite tilting using the Fujikawa approach, their result differs from the one we found. Such difference arises from the alternative ways of dealing with the non-commuting operators involved. Since the calculation is  very different from the  standard case, it turns out that the correct expansion of the regulator required to define the anomaly is highly non-trivial and so it is worthy of the  separate calculation which  we perform.

We will  consider the isotropic case where $v^{j}_a = \delta^{j}_a$, such that  the system to be studied only deals with the tilting of  the cones and it is  given by the Hamiltonian:
\begin{equation}
\label{Hamiltonian}
    \Un{H} = \Un{H}_{-}\oplus\Un{H}_{+},\qquad \Un{H}_{-}(\bld{k}) = -\sigma^0\bld{v}_0\cdot \bld{k}-\pmb{\sigma}\cdot\bld{k}, \qquad \Un{H}_{+}(\bld{k}) = +\sigma^0 \bld{v}_0\cdot \bld{k}+\pmb{\sigma}\cdot\bld{k}.
\end{equation}

Choosing the Weyl basis, we take:

\begin{equation}
    \gamma^0 = \begin{pmatrix}
    0 & 1 \\
    1 & 0 \\
    \end{pmatrix}, \hspace{0.5cm}
    \gamma^a = \begin{pmatrix}
    0 & \sigma^a \\
    -\sigma^a & 0 \\
    \end{pmatrix}, \hspace{0.5cm}
    \gamma^5 = \begin{pmatrix}
    -1 & 0 \\
    0 & 1 \\
    \end{pmatrix},
    \hspace{0.5cm}
    \alpha^a = \gamma^0 \gamma^a= \begin{pmatrix}
    -\sigma^a & 0 \\
     0&  \sigma^a \\
    \end{pmatrix}.
    \label{WEYLBASIS}
\end{equation}

\noindent in Minkowski space. 
Since the matrix $\gamma_5$ will play an important role in this work, we specify its definition together with our convention for the related Levi--Civita symbol $\epsilon_M^{\mu\nu\alpha\beta}$:
\beq
\gamma_5= i \gamma^0 \gamma^1\gamma^2 \gamma^3,  \qquad {\rm tr \, } \gamma_5  \gamma^\mu \, \gamma^\nu\, \gamma^\alpha \, \gamma^\beta= -4i \epsilon_M^{\mu\nu\alpha\beta}, \qquad \epsilon_M^{0123}=+1.
\eeq
On the  basis of (\ref{WEYLBASIS}), the Hamiltonian $\Un{H}$ is: 
\beq
\Un{H}=\begin{pmatrix}
    \Un{H}_+& 0 \\
    0 & \Un{H}_- \\
    \end{pmatrix}= \pmb{\alpha}\cdot \bld{k} +{\bld{v}}_0\cdot \bld{k} \gamma^5.
    \label{HAM1}
\eeq

We then show that  the Lagrangian of Eq.~(\ref{Lagrangian}) reproduces  the Hamiltonian of \mbox{Eq. (\ref{HAM1})} before adding the electromagnetic coupling. We work in momentum space where  $iD_{\mu} \longrightarrow i\partial_{\mu} \longrightarrow -k_{\mu}$. With~this, a~direct calculation from the Lagrangian (\ref{Lagrangian}) yields:
\begin{equation}
\begin{split}
\mathcal{L} & = -\bar{\Psi}\Big[\gamma^{A}k_{A}+\gamma^0\gamma^5\bld{v}_0\cdot\bld{k}\Big]\Psi = -{\Psi}^\dagger \Big[(k_0+\gamma^5\bld{v}^0\cdot\bld{k})+\alpha^a k_a\Big]\Psi
\end{split}, 
\end{equation}
which gives $\Un{H}$ in Eq. (\ref{HAM1}) after the Legendre transformation:
\begin{equation}
\label{Legendre}
    \mathcal{H} = \Pi_0(\partial_0\psi) - \mathcal{L} = -\psi^{\dagger}k_0\psi-\mathcal{L}= \Psi^\dagger \Un{H} \, \Psi,
\end{equation}
is~performed.

\section{The Axial~Anomaly}
\label{AXIALAN}

Before explaining the details, we give a quick overview on how the  axial anomaly 
arises from a general chiral rotation in the fermionic functional for massless fermions, according to the general method of Fujikawa in the path integral approach~\cite{FUJIKAWA,BERTLEMENT}.    The~starting point is: 
\beq
{\mathcal Z}(A_\mu)=\int {\cal D}\bar{\Psi}' {\cal D} \Psi' \exp\Big[ i S(A_\mu \, , \Psi' )\Big]=\int {\cal D}\bar{\Psi}' {\cal D} \Psi' \exp\Big[ i \int d^4 x   \Big(\bar{\Psi}' i \mat{D} \Psi' \Big) \Big],
\label{FERM_FUNC}
\eeq
with $\mat{D}=\big(\gamma^{\mu}D_{\mu}+\gamma^0\gamma^5v^iD_i\big)=\Gamma^\mu (\partial_\mu -ieA_\mu)$.  The~change of integration variables $\Psi(x) \rightarrow \Psi'(x)$ due to the  chiral transformations:
\beq
\Psi'(x)=  e^{i\theta(x)\gamma^5} \Psi(x),\quad 
\bar{\Psi}'(x) = \bar{\Psi}(x) e^{i\theta(x)\gamma^5}, 
\label{QT}
\eeq 
with $ \theta(x)$ arbitrary, leaves  the functional  integral {unchanged} and produces:
\beq
{\mathcal Z}(A_\mu)=\int J(A_\mu )\, {\cal D}\bar{\Psi} {\cal D} \Psi \exp\Big[ i \int d^4 x   \Big(\bar{\Psi} i \mat{D} \Psi-
 J^\mu_5 \,\partial_\mu\theta(x) \Big) \Big].
 \label{ZA_AA}
\eeq 

This  allows the identification of the chiral current as  $ J^\mu_5= \bar{\Psi}(x)\Gamma^\mu \gamma_5 \Psi(x)$ together with the introduction of the  Jacobian  $J(A_\mu)$
required by  the change of variables  ${\cal D}\bar{\Psi}' {\cal D}{\Psi}'= J(A_\mu) \, {\cal D} \bar{\Psi}
 {\cal D} \Psi $. The~further calculation of  $J(A_\mu)$ yields: 
 \beq
 J(A_\mu)=\exp\Big(-i \int d^4 x \, \theta(x) {\cal A}(x)  \Big). 
 \eeq
 
 Since ${\delta {\mathcal Z}(A_\mu)}/{\delta \theta(x)=0}$, it follows that:
\begin{equation}
 \partial_\mu J_5^\mu={\mathcal A}(x),
 \end{equation}
which defines the abelian chiral anomaly ${\mathcal A}(x)$ of the corresponding~current.

The calculation is performed in Euclidean space and we follow the conventions of Ref.~\cite{FUJIKAWA}. We perform a Wick rotation:
\begin{equation}
\begin{split}
    x^0 & \longrightarrow -ix^4,  \qquad 
    x_0  \longrightarrow ix_4, \hspace{0.5cm} x_4\in \mathbb{R}, \quad \gamma^0  \longrightarrow -i\gamma^4,
\end{split}
\label{WR}
\end{equation}
{which applies to all tensorial indices}. This leaves  us  with the Euclidean the metric $\eta_{\mu\nu} = \mathrm{diag}(-1,-1,-1,-1)$ and the Lorentz indices $\mu$  now run from $1$ to $4$. 
The Euclidean Dirac matrices satisfy:
\beq
\{ \gamma^\mu, \gamma^\nu \}=2\eta^{\mu\nu},\qquad (\gamma^\mu)^\dagger=- \gamma^\mu
\eeq
and the pseudoscalar $\gamma^{5}$ retains its definition in Minkowski space:
\begin{equation}
\label{gamma5}
\gamma^5 = i\gamma^0\gamma^1\gamma^2\gamma^3 = \gamma^4\gamma^1\gamma^2\gamma^3 = - \gamma^1\gamma^2\gamma^3\gamma^4,
\end{equation}
with the basic properties:
\beq
{\rm tr\, } \gamma_5=0, \qquad  {\rm tr\, } \gamma_5 \gamma^\mu \gamma^\nu=0, \qquad  
{\rm tr\, } \gamma_5 \gamma^\mu \gamma^\nu \gamma^\alpha \gamma^\beta= -4\epsilon_E^{\mu\nu\alpha \beta}, \qquad \epsilon_E^{1234}=+1.
\label{TRGAMMA5}
\eeq

Since $v_0^i$ is contracted  with the space indices of the covariant derivatives $D_i$, we consider these parameters as the space components of a vector which are not affected by the Wick rotation. Thus, we have:
\begin{equation}
    {v}_0^i \longrightarrow {v}_4^i, \qquad v_4^i\in \mathbb{R}.
    \label{V0E}
\end{equation}
With the above choices, the Euclidean Dirac operator: 
\beq
\mat{D}_{E}= \gamma^{\mu}D_{\mu}-i\gamma^4\gamma^5 v_4^kD_k,
\eeq
is Hermitian under the scalar product $(\psi, \mat{D}_E \, \chi)=\int d^4 x \psi^\dagger(x) \mat{D}_E \, \chi(x) $. The~Euclidean action is:
\beq
S_E=\int d^4 x \, {\cal L}_E, \quad \quad   {\cal L}_E= \bar{\Psi}(x)\, i    \mat{D}_E \, \Psi(x), 
\label{Lagrangian_E}
\eeq
where $\bar{\Psi}(x)$ and $\Psi(x)$ are now independent fields. When going back to the Minkowski space, we set $\Psi \rightarrow \Psi_M$,  $\bar{\Psi} \rightarrow (\Psi_M)^\dagger \gamma^0$, $d^4 x \rightarrow i (d^4 x)_M$ 
and $P^\mu \eta_{\mu \nu} Q^\nu\rightarrow  (P^\mu \eta_{\mu \nu} Q^\nu)_M\, $, in addition to using the relations in Eq. (\ref{WR}).

\subsection{The Modified Axial~Current}
\label{MAXIAL}

Our Lagrangian (\ref{Lagrangian_E}) is trivially invariant under the  global transformations:
\begin{equation}
    \Psi(x) \longrightarrow e^{i\theta}\Psi(x),\quad
    \bar{\Psi}(x) \longrightarrow \bar{\Psi}(x)e^{-i\theta},
\end{equation}
which lead to the charge  Noether current:
\begin{equation}
    J^{\mu} = \bar{\Psi}\gamma^{\mu}\Psi-i\bar{\Psi}\gamma^4\gamma^5v_4^k\delta^{\mu}_k\Psi,
\end{equation}
such that $\partial_\mu J^\mu=0.$ The invariance of the Lagrangian can be extended to transformations including a local parameter $\theta(x)$, provided we demand that $\quad A_{\mu}\longrightarrow A_{\mu}-\partial_{\mu}\theta(x)/e$. Now, the conservation of $J^\mu$ is a consequence of gauge invariance, and must still hold in the quantum  version of the theory. In~other words, no anomalies are acceptable  for this~current. 

The massless classical Lagrangian (\ref{Lagrangian_E}) has an additional symmetry corresponding to the chiral  transformations
(\ref{QT})
with the constant parameter $\theta$. The~associated  Noether current, which is conserved due to the classical equations of motion, can be readily identified by an arbitrary infinitesimal variation of the fields: 
\beq
\delta {\bar \Psi}(x)=i \delta \theta(x){\bar \Psi}\gamma^5, \qquad \delta { \Psi}(x)=i \delta \theta(x) \gamma^5 \Psi(x)
\label{INF_QT}
\eeq
resulting when promoting $\delta \theta$ to a local parameter. The~equations of motions yield:
\beq
0=\delta S_E= \delta \int d^4 x \, \delta {\cal L}_E= -\int d^4 x \,(\partial_{\mu}\delta \theta) \Big[ \bar{\Psi}\gamma^{\mu}\gamma^5\Psi-i (v_4^k\delta^\mu_k)\bar{\Psi}\gamma^4\Psi \Big],
\eeq
from where we read  the  conservation of the modified classical axial current:
\beq
J^\mu_5=\bar{\Psi}\gamma^{\mu}\gamma^5\Psi-i (v_4^k\delta^\mu_k)\bar{\Psi}\gamma^4\Psi,
\label{j5}
\eeq
since $\delta \theta(x)$ is arbitrary.
In quantum electrodynamics, it is not possible to maintain the conservation of both $J^\mu$ and $J^\mu_5$, and one must choose the conservation of the electric charge to preserve gauge invariance. This leads to the presence of the chiral~anomaly.

\subsection{The Fujikawa~Method}
\label{FUJIM}

{As we previously mentioned, the~path integral calculation of the chiral anomaly~\cite{FUJIKAWA}} tells us that its origin  is found in  the non-trivial Jacobian, arising when performing the axial transformation. {Using this method},  we will compute the  axial  anomaly {originating  from}  our modified Lagrangian (\ref{Lagrangian_E}). We start  from the   Euclidean path integral: 
\beq
{\mathcal Z}_E(A_\mu)=\int {\cal D}\bar{\Psi}_E {\cal D} \Psi_E \exp S_E =\int {\cal D}\bar{\Psi}_E {\cal D} \Psi_E \exp\Big[  \int (d^4 x)_E   \Big(\bar{\Psi}_E i \mat{D}_E \Psi_E \Big) \Big],
\label{BASIC_EI}
\eeq
where we suppress the subindices $E$ in the following.
Recalling that our modified Dirac operator $\mat{D}$ is Hermitian, with~eigenfunctions $\varphi_n(x)$ such that $\mat{D}\varphi_n= \lambda_n \varphi_n$, we follow the standard steps in the calculation of the Jacobian by 
expanding the fields in terms of the orthonormal and complete eigenvectors $\varphi_n(x)$. That is:
\begin{equation}
    \Psi(x) = \sum_n a_n\varphi_n(x),\quad
    \bar{\Psi}(x) = \sum_n \bar{b}_n\varphi^{\dagger}_n(x),
\end{equation}
and similarly for ${\bar {\Psi}'}$ and  $\Psi'$.  Let us recall that the fields are Grassmann numbers and so are the coefficients 
$a_n, a'_n$ and ${\bar b}_n, {\bar {b}'}_n $. The~infinitesimal axial transformation (\ref{INF_QT}) yields the~relation:
\beq
a'_n= \Big(\delta_{n,m}+i\int \mathrm{d}^4x\varphi^{\dagger}_n(x)\delta \theta(x)\gamma^5\varphi_m(x)\Big) a_n \equiv T_{mn} a_n,
\eeq 
with a similar expression for ${\bar b}'_n$ in terms of ${\bar b}_n$. The~above equation introduces the  matrix $T=I+\delta t$ with:
\beq
{\delta t}_{mn}=i\int \mathrm{d}^4x\varphi^{\dagger}_n(x)\, \delta \theta(x)\, \gamma^5\varphi_m(x).
\eeq

 The Grassmann character of the expansion coefficients leads to:
\begin{equation}
\begin{split}
    \mathcal{D}\bar{\Psi'}\mathcal{D}\Psi' = \lim_{N \longrightarrow \infty}\prod^{N}_{n=1} \mathrm{d}\bar{b'}_n\mathrm{d}a'_n = (\det T)^{-2} & \lim_{N \longrightarrow \infty} \prod^{N}_{n=1} \mathrm{d}\bar{b}_n\mathrm{d}a_n =  (\det T)^{-2} \, 
    \mathcal{D}\bar{\Psi}\mathcal{D}\Psi,
\end{split}
\end{equation}
which identifies $J=(\det T)^{-2}$ as the Jacobian of the transformation. As~$\det(M) = \exp\big[\Tr(\ln(M))\big]$, we have:
\barr
&&J_{\delta \theta}=(\det T)^{-2}= \exp\big[-2\Tr(\ln(I+\delta t))\big]=
\exp\big[-2\, \Tr(\delta t)\big],\nonumber \\
&&J_{\delta \theta}=  \exp\Big[-2i\sum_n\int \mathrm{d}^4x\varphi^{\dagger}_n(x)\delta \theta(x)\gamma^5\varphi_n(x)\Big]= 
 \exp\Big[-2i \Tr
 \, \big( \delta \theta \,   \gamma^5 \big) \Big],
\earr
to the first order in $ \delta \theta(x)$. Here, the trace $\Tr{} $ is taken in the matrix space ($\sum_n$) as well as  in the coordinate space ($\int \mathrm{d}^4x $).
Nevertheless, the~series in the exponential diverges, and~so it must be regularized. We need to preserve  invariance under the gauge  transformations: 
\beq
{\varphi'}_n (x)= e^{i \Lambda(x)} \varphi_n (x),  \quad A'_\mu
= A_\mu-\frac{1}{e}\partial_\mu \Lambda(x) 
\eeq
which induce the transformation $\mat{D}e^{i \Lambda (x)}=e^{i \Lambda (x)} \mat{D} \, $ such that $\varphi^\dagger_n (x)F(\mat{D}) \varphi_n (x)$ is gauge invariant. To~ensure convergence, we choose $F(\mat{D})= \exp[-\mat{D}^2/M^2]$ with $M \rightarrow \infty$ at the end of the~calculation.

It is convenient to define:
\begin{equation}
    \alpha(x) \equiv \lim_{M\longrightarrow \infty}\sum_n\varphi^{\dagger}_n(x)\gamma^5e^{-\mat{D}^2/M^2}\varphi_n(x),
\end{equation}
such that the regularized Jacobian is written as 
\beq
J(A_\mu)=\exp \Bigg[-2i \int \mathrm{d}^4x \, \delta \theta(x) \,  \alpha(x)  \Bigg].
\label{REG_JAC}
\eeq
This expression for the regularized Jacobian, together with the change in the Euclidean action (\ref{BASIC_EI}) due to the chiral transformations (\ref{INF_QT}), enters in the functional integral by modifying the Euclidean Lagrangian (\ref{Lagrangian_E}) as 
\beq
{\cal L}_E \rightarrow  {\cal L}_E  +\delta \theta(x) \Big( \partial_\mu J_5^\mu - 2i\, \alpha(x) \Big), 
\label{MOD_LE}
\eeq
where $\alpha(x)$ depends on the external electromagnetic field and $J^\mu_5$ is given by Eq. (\ref{j5}). Let us emphasize that to calculate the chiral anomaly, it is sufficient to { consider} only the infinitesimal transformations (\ref{INF_QT}).

Having regularized the Jacobian, we take a convenient change of basis, from~$\{\varphi_n(x)\}$ to $\{e^{ik\cdot x}\}$, so~that:
\begin{equation}
    \alpha(x) = \lim_{M\longrightarrow \infty}{\rm tr}\int \frac{\mathrm{d}^4k}{(2\pi)^4}e^{-ik\cdot x}\gamma^5e^{-\mat{D}^2/M^2}e^{ik\cdot x}.
    \label{alpha}
\end{equation}

Here, ${\rm tr}$ refers to the trace in the gamma matrix space. We then calculate $\mat{D}^2 $ and organize the resulting terms to evaluate $\alpha(x)$. Using the relation $[D_{\mu},D_{\nu}] = -ieF_{\mu\nu}$, we~obtain:
\begin{equation}
\begin{split}
    \mat{D}^2 & = D_{\mu}D^{\mu}-\frac{ie}{4}[\gamma^{\mu},\gamma^{\nu}]F_{\mu\nu}-e\gamma^{\mu}\gamma^4\gamma^5v_4^jF_{\mu j}-2i\gamma^a\gamma^4\gamma^5v_4^jD_jD_a-(v_4^iD_i)^2. \\
\end{split}
\end{equation}

As $[D_{\mu},e^{ik\cdot x}] = ik_{\mu}e^{ik\cdot x}$, we can remove the exponentials in  Eq. (\ref{alpha}) by shifting $D_\mu \rightarrow D_\mu+i k_\mu$. This leads to:
\begin{equation}
\begin{split}
    \alpha(x) = \lim_{M\longrightarrow \infty}{\rm tr} \, \gamma^5 \int \frac{\mathrm{d}^4k}{(2\pi)^4}& \exp \frac{-1}{M^2}\bigg[ (ik_{\mu}+D_{\mu})^2-\frac{ie}{4}[\gamma^{\mu},\gamma^{\nu}]F_{\mu\nu}-e\gamma^{\mu}\gamma^4\gamma^5v_4^iF_{\mu i} \\
    & -2i\gamma^i\gamma^4\gamma^5v_4^j(ik_j+D_j)(ik_i+D_i)-(v_4^j(ik_j+D_j))^2\bigg]. \\
\end{split}
\end{equation}

Scaling the momentum integral $k_{\mu}\longrightarrow Mk_{\mu}$, and~defining $\lambda=1/M$, we obtain the final expression  :
\begin{equation}
\label{ANOMALY}
\begin{split}
    \alpha(x) & = \lim_{\lambda\longrightarrow 0} \int \frac{\mathrm{d}^4k}{(2\pi)^4}
    e^{-k_{\mu}k_{\mu}-(v^4_ak_a)^2}\Bigg( \frac{1}{\lambda^4} {\rm tr} \,  {\gamma^5} e^{\mat{A}+\lambda \mat{B}}\Bigg).
\end{split}
\end{equation}
with the identifications:
\begin{equation}
\mat{A} = -2i\gamma^i\gamma^4\gamma^5v_4^jk_jk_i,
    \qquad  \mat{B} = \alpha +\beta_i\gamma^i\gamma^4\gamma^5+\frac{ie}{2}\lambda(\gamma^{\mu}\gamma^{\nu}F_{\mu\nu}-2i\gamma^{\mu}\gamma^4\gamma^5v_4^jF_{\mu j}),
\end{equation}
and  the further splitting:
\begin{equation}
    \alpha = 2i(v_4^ik_i v_4^jD_j-k_{\mu}D^{\mu})+\lambda((v_4^jD_j)^2-D_{\mu}D^{\mu}), \quad \beta_i = -2v_4^j(k_jD_i+k_iD_j)+2i\lambda v_4^jD_jD_i.
\end{equation}

As the operators $\mat{A}$ and $\mat{B}$ do not commute, the~factorization $e^{\mat{A}+\lambda \mat{B}}=e^\mat{A} \, e^{\lambda \mat{B}} $ followed by an expansion of $e^{\lambda \mat{B}}$ in a Taylor series of $\lambda \mat{B}$ is not allowed. We must therefore take the complete expansion:
\begin{equation}
\label{LV6}
e^{\mat{A}+\lambda \mat{B}}=\sum_{n=0}^\infty \frac{1}{n!}(\mat{A}+\lambda \mat{B})^n.
\end{equation}

This includes terms to all orders of $v_4^i$, even when focusing on those proportional to
$\lambda^4$, which are the only ones  which  contribute to $\alpha(x)$. 

In this work, we restrict ourselves to calculate  the contributions to $\alpha(x)$ only to the first and second order in $v_4^i$. This means that  we are considering  $v_4^i$ as a small parameter, which limits the range of our results in the context of further applications to  Weyl semimetals. With~this assumption, we need to  identify the required powers of $v_4^i$ in $\mat{A}$ and $\mat{B}$. To~simplify the notation, we write $v_4^i=v^i$  in what follows:
\begin{eqnarray}
&&\mat{A} =v^j\;M_{j},\;\;\;\;M_{j}=-2i\left( \gamma ^{i}\gamma ^{4}\gamma
^{5}\right) k_{j}k_{i}, \label{38}\\
&&
\mat{B}=N_{0}+v^jN_{j}+v^iv^jN_{ij}, \label{39}\\
&&
N_{0} =-2ik_{\mu }D^{\mu }+\lambda \left( \frac{ie}{2}\gamma ^{\mu }\gamma
^{\nu }F_{\mu \nu }-D_{\mu }D^{\mu }\right),  \notag \\
&&N_{j} =-2(k_{j}D_{i}+k_{i}D_{j})\gamma ^{i}\gamma ^{4}\gamma ^{5}+\lambda
\left( 2iD_{j}D_{i}\gamma ^{i}\gamma ^{4}\gamma ^{5}+e\gamma ^{\mu }\gamma
^{4}\gamma ^{5}F_{\mu j}\right),  \notag \\
&& N_{ij} =2ik_{i}D_{j}+\lambda D_{i}D_{j}.
\label{40}
\end{eqnarray}

Since any contribution to $\alpha(x)$ arises from the term 
proportional to $\lambda^4$ in the expansion of the exponential (\ref{LV6}), we still have to explicitly write the $\lambda$-dependence of the above coefficients. This yields the additional splitting:
\begin{eqnarray}
N_{0} =
 N_{00}+\lambda N_{01}, \qquad  
N_{j} 
= N_{j0}+\lambda N_{j1}, \qquad 
N_{ij} =N_{ij0}+\lambda N_{ij1},
\label{LV10}
\end{eqnarray}
with:
\begin{eqnarray}
N_{00} &=&-2ik_{\mu }D^{\mu },\;\;\;N_{01}=\left( \frac{ie}{2}\gamma ^{\mu
}\gamma ^{\nu }F_{\mu \nu }-D_{\mu }D^{\mu }\right), \\
N_{j0} &=&-2(k_{j}D_{i}+k_{i}D_{j})\left( \gamma ^{i}\gamma ^{4}\gamma
^{5}\right) , \\
N_{j1} &=&\left[ e\gamma ^{\mu }\gamma ^{4}\gamma ^{5}F_{\mu
j}+2iD_{j}D_{i}\left( \gamma ^{i}\gamma ^{4}\gamma ^{5}\right) \right], \\
\;\;N_{ij0} &=&2ik_{i}D_{j},\;\;\;\;\;\;N_{ij1}=D_{i}D_{j}.
\label{FULL_EXP}
\end{eqnarray}

Let us observe that all the operators in the previous four equations commute with $\gamma_5$.  
Since the binomial expansion in Eq. (\ref{LV6}) will produce  multiple permutations of the different coordinate operators ($D_{\mu}$ and $F_{\mu\nu}$), it would be tempting to use the cyclic property of the trace in coordinate space to move operators from the far right to the far left of some expressions and therefore  simplify the proliferation of non-commutative contributions. Nevertheless, this would not be useful in the case of derivatives because of the term $\delta \theta (x)$ inside the spacetime  trace. Suppose we have a term like  ${\rm Tr} \, \delta \theta(x) \gamma_5 
D_\mu F_{\alpha \beta} D_\nu $, then the cyclic identity will yield  ${\rm Tr} \,  D_\nu \, \delta \theta(x) \gamma_5 
D_\mu F_{\alpha \beta} $  which is different from 
${\rm Tr} \, \delta \theta(x) \gamma_5 
D_\nu D_\mu F_{\alpha \beta}  $  since $[D_\nu, \delta \theta (x)]=-(\partial_\nu \delta \theta (x)) \neq 0$. For~this reason,  we only use the cyclic identity of the trace ${\rm tr}$ in matrix space in our~calculations.

\section{The First Order~Expansion}
\label{First_Order}

We look for the first order contribution  in $v^i$ to the anomaly $\alpha(x)$ defined in Eq. (\ref{ANOMALY}). To~take into account the non-commuting operators that are involved, we start by rewriting the term $(\mat{A}+\lambda \mat{B})^n$ to be expanded so that the dependence upon $v_a$ is made explicit:

\barr
&&\frac{1}{n!}(\mat{A} +\lambda \mat{B})^n= \frac{1}{n!}(A+ B_i v^i + v^iv^j C_{ij})^n,\nonumber \\
&& A=\lambda N_0, \qquad B_i= M_i + \lambda N_i, \qquad  C_{ij}= \lambda N_{ij}.
\label{EXP_V}
\earr

The coefficients $N_0,\, M_i,\, N_i, \, N_{ij}$ were previously defined in Eqs. (\ref{38}) and (\ref{LV10}).  As~previously emphasized, we still have to make explicit the $\lambda$-dependence of the expansion and only focus on the $\lambda^4$ contribution. Furthermore, we recall that $\gamma^5$ commutes with all the remaining operators. The~contribution to each order $n$ is denoted by $a_n$. In~this way, when looking for the linear contribution in $v^i$, it is sufficient to consider the terms:
\beq
\frac{1}{n!}(\mat{A} +\lambda \mat{
B})^n\rightarrow \frac{1}{n!}\lf A^{n-1}B_i  \rf  v^i\equiv a_n, 
\eeq
where $\lf A^{n-1}B_i  \rf$ is a short-hand notation for all possible ordering arising in a given product of operators which do not commute. For~example, in~this case, the coefficients  $a_n$ are:
\barr
&&  a_{2}=\frac{1}{2!} \lf A, B_i \rf v^i=\frac{1}{2!}\left( AB_{i}+B_{i}A\right) v^i, \nonumber  \\
&&a_{3} = \frac{1}{3!} \lf A^2, B_i \rf v^i=\frac{1}{3!}\left( AB_{i}A+B_{i}A^{2}+A^{2}B_{i}\right) v^i, \nonumber \\
&&a_{4} =\frac{1}{4!} \lf A^3,B_i \rf v^i=\frac{1}{4!}\left(
AB_{i}A^{2}+B_{i}A^{3}+A^{2}B_{i}A+A^{3}B_{i}\right) v^i,
\nonumber \\
&&
a_{5}=\frac{1}{5!} \lf A^4,B_i\rf v^i=\frac{1}{5!}\left(
A^{4}B_{i}+AB_{i}A^{3}+A^{2}B_{i}A^{2}+A^{3}B_{i}A+B_{i}A^{4}\right) v^i.
\label{DEFSYMBOL}
\earr

Since $A=\lambda N_0$, all contributions $a_n$ with $  n \geq 6$
will yield a zero~result. 

Now, we make explicit the $\lambda$-dependence in order to extract the $\lambda^4$ terms. To~this end, we use Eq. (\ref{EXP_V}) together with the further expansion in powers of $\lambda$ given in \mbox{Eq. (\ref{LV10})}. This yields:
\begin{eqnarray}
&& a_{2} =\frac{1}{2!}\lambda \lf \left( N_{00}+\lambda N_{01}\right), \left(
M_{i}+\lambda N_{i0}+\lambda ^{2}N_{i1}\right)\rf_2
v^i, \\
&& a_{3} =\frac{1}{3!}
\lf
\lambda ^{2}\left( N_{00}+\lambda N_{01}\right)
^{2}, \left
( M_{i}+\lambda N_{i0}+\lambda ^{2}N_{i1}\right)\rf_3
v^i, \\
&&a_{4} =\frac{1}{4!}\lf
\lambda ^{3}\left( N_{00}+\lambda N_{01}\right)
^{3},
\left( M_{i}+\lambda N_{i0}+\lambda ^{2}N_{i1}\right)
\rf_4
 v^i , \\
&& a_{5} =\frac{1}{5!}\lf
\lambda ^{4}\left( N_{00}+\lambda N_{01}\right)
^{4},
\left( M_{i}+\lambda N_{i0}+\lambda ^{2}N_{i1}\right)\rf_5
v^i,
\end{eqnarray}
from where we extract the coefficients of $\lambda^4$.
 The result is: 
\barr
&&\frac{{a}_{2}}{\lambda ^{4}}=\frac{1}{2!}
\lf
N_{01},
N_{i1}\rf_2 
v^i
,
\label{65N}\\
&&
\frac{{a}_{3}}{\lambda ^{4}}=\frac{1}{3!}
\Big
[\lf
 N_{01}^{2}, 
M_{i}\rf_3
 +\lf
N_{00},
N_{01},
N_{i0}\rf_6
+\lf
N_{00}^{2},
N_{i1}\rf_3
\Big]
v^i, \label{66N}\\
&&
\frac{{a}_{4}}{\lambda ^{4}}=\frac{1}{4!}
\Big[
\lf
N_{00}^{2}, 
N_{01},
M_{i}\rf_{24}
+\lf
N_{00}^{3},
N_{i0}\rf_{4}
\Big] v^i,\label{67N}\\
&&
\frac{{a}_{5}}{\lambda ^{4}}=\frac{1}{5!} \lf N_{00}^{4}, M_{i}\rf_5 v^i.
\label{68N}
\earr

The notation is $\lf P, Q,\dots R \rf_n$, where $n$ denotes the total number of non-commuting products inside the bracket $\lf \dots \rf $.
Now, we take the trace in matrix space including $\gamma^5$,
obtaining:
\barr
&&\hspace{-1cm} {\rm tr} \Big(  \gamma^{5} \lf N_{01}, N_{j1}\rf_2 \Big) = 8ie^{2}F^{4\mu }F_{\mu j}+4e\left\{ F_{4j},D_{\mu }D^{\mu }\right\}
-8e\;\left\{ D_{j}D_{i},F^{4i}\right\},
\\
&& \hspace{-1cm}
{\rm tr} \Big( \gamma ^{5} \lf N_{01}^{2}, M_{j}\rf_3\Big) = 24e(k_{j}k_{i})\left\{ D_{\alpha }D^{\alpha },F_{4i}\right\} , \\
&& \hspace{-1cm} {\rm tr}\Big(\gamma _{5\;}\left\lfloor N_{00}^{2},N_{j1}\right\rfloor _{3}\Big)=16e%
\left[ \left\{ \left( k_\mu D^{\mu }\right) ^{2},\,F_{4j}\right\} +\left(
k_{\mu }D^{\mu }\right) F_{4j}\left( k_{\mu }D^{\mu }\right) \right], \\
&& \hspace{-1cm}
{\rm tr} \Big( \gamma ^{5}\lf N_{00}, N_{01}, N_{j0}
\rf_6 \Big)
= -16e\left( k_{\alpha }D^{\alpha }\right) \left\{
F_{4m},\;(k_{j}D_{m}+k_{m}D_{j})\right\} \nonumber    \\
&& \hspace{3cm} -16e(k_{j}D_{m}+k_{m}D_{j})\left\{ k_{\alpha }D^{\alpha },F_{4m}\right\} \nonumber \\
&& \hspace{3cm} -16eF_{4m}\left\{ k_{\alpha }D^{\alpha },\;(k_{j}D_{m}+k_{m}D_{j})\right\}, \\
&&\hspace{-1cm}
{\rm tr} \Big( \gamma ^{5}\lf
N_{00}^{2}, N_{01}, M_{j} \rf_{24} \Big)=-128ek_{j}k_{i}\left[ \left\{ \left( k_{\mu }D^{\mu }\right)
^{2},\;F_{4i}\right\} +\left( k_{\mu }D^{\mu }\right) F_{4i}\left( k_{\alpha
}D^{\alpha }\right) \right],  \label{63}\\
&& \hspace{-1cm}
{\rm tr} \Big(\gamma ^{5} \lf N_{00}^{3}, N_{j0} \rf_4\Big)=0 , \qquad  {\rm tr} \Big(\gamma ^{5}\lf N_{00}^{4}, M_{j}\rf_4\Big)=0.
\earr

In the above equations, we factored out the term $v^j$ and $\{ P, Q\}=PQ + QP$ denotes the anticommutator of $A$ and $B$.

We then perform the integrals with respect to the momentum using the relations:
\begin{eqnarray}
\label{integrals}
&&\int d^{4}ke^{-k_{\mu }k_{\mu }}\;()\; \rightarrow \;\;\pi ^{2}(), \qquad 
\int d^{4}ke^{-k_{\mu }k_{\mu }}k_{\alpha }k_{\beta }() =\frac{\pi ^{2}}{2}%
\delta _{\alpha \beta }(),  \nonumber \\
&&\int d^{4}ke^{-k_{\mu }k_{\mu }}k_{\alpha }k_{\beta }k_{\mu }k_{\nu }() =
\frac{\pi ^{2}}{4}\left( \delta _{\alpha \beta }\delta _{\mu \nu }+\delta
_{\alpha \mu }\delta _{\beta \nu }+\delta _{\alpha \nu }\delta _{\beta \mu
}\right) ().
\end{eqnarray}
Going back to the notation in Eqs. (\ref{65N})--(\ref{68N}),
we obtain:
\begin{eqnarray}
&&\hspace{-1.5cm}\frac{1}{\pi ^{2}\lambda ^{4}}\int d^{4}ke^{-k_{\mu }k_{\mu }}{\rm tr}(\gamma ^{5}%
{a}_{2})_{j}= 4ie^{2}F_{4\mu }F_{\mu
j}+2e \{D^{2}, F_{4j} \}-4e\left\{ F_{4i},\;D_{j}D_{i}\right\}, 
\end{eqnarray}%
\begin{eqnarray}
&&\hspace{-1.5cm} \frac{1}{\pi ^{2}\lambda ^{4}}\int d^{4}ke^{-k_{\mu }k_{\mu }}{\rm tr}\left(
\gamma ^{5}{a}_{3}\right) _{j} =-2e\left\{ D^{2},F_{4j}\right\} -%
\frac{4}{3}e\left[ \left\{ D^{2},F_{4j}\right\} +D^{\mu }F_{4j}D_{\mu }%
\right]   \nonumber \\
&& \hspace{3cm}+\frac{8}{3}e\Big( D_{j}F^{4i}D_{i}+D_{i}F^{4i}D_{j}+\left\{
F^{4i},\left\{ D_{j},\;D_{i}\right\} \right\} \Big), 
\end{eqnarray}%
\begin{eqnarray}
&&\hspace{-1.5cm} \frac{1}{\pi ^{2}\lambda ^{4}}\int d^{4}ke^{-k_{\mu }k_{\mu }}{\rm tr}\left(
\gamma ^{5}{a}_{4}\right) _{j} =+\frac{4}{3}e\Big( \left\{
D^{2},\;F_{4j}\right\} +D^{\mu }F_{4j}D_{\mu }\Big)   \nonumber \\
&&\hspace{3cm}-\frac{4}{3}e\Big( \left\{ \left\{ D_{j},\;D_{i}\right\}
,\;F_{4i}\right\} +D_{j}F_{4i}D_{i}+D_{i}F_{4i}D_{j}\Big). 
\end{eqnarray}

Here, $D^2=D_\mu D^\mu$. In~this notation, we write the  contribution $\alpha^{(1)}(x)$ to the anomaly, which is linear  in $v^j$, as~\beq
(4 \pi)^2 \, \alpha^{(1)}(x)= \frac{1}{\pi ^2 \lambda^4}\int d^4 k \, e^{-k_\mu k_\mu} {\rm tr} (\gamma_5 {a}_j)  \,  v^j, \qquad {a}_j= ({a}_{2}+{a}_{3}+{a}_{4})_j.
\label{SFANOM} 
\eeq

The direct combination of the previous equations in (\ref{SFANOM}) cancels the $D^2$ terms and provides  commutators of covariant derivatives  which can be traded by the corresponding field strength. We are left with:
\begin{eqnarray}
&&\hspace{-1cm} (4 \pi)^2 \, \alpha^{(1)}(x)= -i\frac{4}{3}e^{2}F_{4i}\;F_{ij} v^j
+\frac{4}{3}e\left(
D_{i}F_{4i}D_{j}-F^{4i}D_{i}D_{j}+D_{j}F_{4i}D_{i}-D_{i}D_{j}F_{4i}\right)v^j.
\label{80N}
\end{eqnarray}%

In order to show the gauge invariance of our result, we rearrange the terms with covariant derivatives in the following way:
\begin{eqnarray}
D_{i}F_{4i}D_{j}-F_{4i}D_{i}D_{j} &=&D_{i}F_{4i}D_{j}-\left( \left[
F_{4i},D_{i}\right] +D_{i}F_{4i}\right) D_{j}=
\left( \partial _{i}F^{4i}\right) D_{j}, 
\end{eqnarray}%
\begin{equation}
D_{j} F_{4i}D_{i} -D_{i}D_j F_{4i}=D_{j}[ F_{4i}, D_i] +\left( D_{j}D_{i}-D_{i}D_{j}\right) F_{4i}=-D_j(\partial_i F_{4i})-ieF_{ji}F_{4i}.
\end{equation}%

Substituting in Eq. (\ref{80N}) yields: \begin{eqnarray}
&&(4 \pi)^2 \, \alpha^{(1)}(x) =-i\frac{4}{3}e^{2}F_{4i}\;F_{ij} v^j
+\frac{4}{3}e\Big(
\left( \partial _{i}F^{4i}\right) D_{j} -D_j(\partial_i F_{4i})-ieF_{ji}F_{4i}\Big)v^j\nonumber \\
&&\hspace{2.5cm}=
\frac{4}{3}e v^j\Big(
\left( \partial _{i}F^{4i}\right) D_{j} -D_j(\partial_i F_{4i})\Big)\nonumber \\&=& \frac{4}{3}e v^j\Big(
\left( \partial _{i}F^{4i}\right) (-ieA_j) -(\partial_j\partial_i F_{4i})+(ieA_j )(\partial_i F_{4i})\Big)=-\frac{4}{3}e v^j \partial_j\partial_i F_{4i}.
\end{eqnarray}%

Nevertheless, we can discard  this  unexpected  gauge invariant term, recalling that a redefinition of the current is allowed provided the value of the charge remains unchanged. In~our case, we show
that the  contribution of $\alpha^{(1)}(x)$ to the chiral charge is zero. Returning to the Minkowski space, the~additional term in the divergence of the axial current is  $\partial^\mu J^{(1)\mu}_{5} \sim ev^j\partial_j \partial_\mu F^{\mu 0}=\partial_\mu (ev^j\partial_j F^{\mu 0})$,   which yields the extra term $J^{(1) \mu}_{5}= ev^j\partial_j F^{\mu 0}$ in the axial current. The~corresponding charge: 
\barr
{Q_5}^{(1)}=\int d^3 x \, {{J_5}^{(1)0}}
\label{ACH}
\earr
is identically zero because $F^{00}=0$. 
Thus, we find 
 a null contribution to the anomaly at first order in $v^i$.

\section{The Second Order~Expansion}
\label{Second_Order}

Since in  flat space  there is no difference among spacetime and Lorentz indices, we include $a,b, \dots$ among the space indices in the following. 
Having obtained that there are no linear contributions of $v^a$ to the anomaly, we now focus on the quadratic ones.   They are included in: 
\barr
\frac{1}{n!}(\mat{A}+\lambda \mat{B})^{n} &\rightarrow & \frac{1}{n!} \lf
(\lambda N_0)^{n-2}, (M_a+ \lambda N_a), (M_b + \lambda N_b)\rf_{\left( _{2}^{n}\right)}
  v^{a}v^{b} \nonumber   \\
&& + \frac{1}{n!} \lf (\lambda N_0)^{n-1}, (\lambda N_{ab})\rf_n  v^{a}v^{b}   \label{BNmasCN} \\
&& \equiv  b_n + c_n,
\earr
where $b_n$  and $ c_n$ denote the first and second term in the right-hand side of Eq. (\ref{BNmasCN}), respectively. Let us also keep in mind the definition of the symbol $\lf \dots \rf$  given in Eq. (\ref{DEFSYMBOL}).

The only relevant contributions ($\lambda$ of order  less than 5) are $b_2$, $b_3$, $b_4$, $b_5$, $b_6$ and $c_1$, $c_2$, $c_3$, $c_4$. This means:
\begin{eqnarray}
&& b_2 = \frac12 \lf (M_a+ \lambda N_a)(M_b + \lambda N_b)\rf_1 v^{a}v^{b}, \nonumber \\ 
&& b_3 = \frac16\lf  (\lambda N_0), (M_a+ \lambda N_a)(M_b + \lambda N_b)\rf_3 v^{a}v^{b}, \notag \\
&& b_4 = \frac{1}{24}\lf 
(\lambda N_0)^{2}, (M_a+ \lambda N_a) (M_b + \lambda N_b)\rf_6 v^{a}v^{b}, \notag \\
&& b_5 = \frac{1}{120}\lf 
(\lambda N_0)^{3}, (M_a+ \lambda N_a) (M_b + \lambda N_b)\rf_{10} v^{a}v^{b}, \notag \\
&& b_6 = \frac{1}{720}\lf 
(\lambda N_0)^{4}, (M_a+ \lambda N_a) (M_b + \lambda N_b)\rf_{15} v^{a}v^{b},
\end{eqnarray}
and:
\begin{eqnarray}
&& c_1 = \lf \lambda N_{ab}\rf_1 v^{a}v^{b}, \qquad
c_2 = \frac{1}{2}\lf (\lambda N_0), (\lambda N_{ab})\rf_2 v^{a}v^{b}, \quad
 \notag \\
&& c_3 = \frac{1}{6}\lf (\lambda N_0)^{2}, (\lambda N_{ab})
\rf_3 v^{a}v^{b}, \qquad c_4 = \frac{1}{24}\lf (\lambda N_0)^{3}, (\lambda N_{ab})\rf_4 v^{a}v^{b}.
\end{eqnarray}

In our notation, no permutations are made between the operators $(M_a+\lambda N_a)$ and $(M_b+\lambda N_b)$ as they are the same when contracted with $v_a$ and $v_b$. To~emphasize  this, we introduced
the further convention that only the terms separated with commas in $\lf \dots \rf$ are subjected to permutations.  For~clarity, the~subindex in $\lf \dots \rf$  indicates the number of permutations in the bracket, which is given by a multinomial  coefficient. Making explicit the full dependence in $\lambda$, we have:

\begin{eqnarray}
&& b_2 = \frac12 \lf(M_a+ \lambda N_{a0}+\lambda^2 N_{a1}) (M_b + \lambda N_{b0}+\lambda^2N_{b1})\rf_1 v^{a}v^{b}, \notag \\
&& b_3 = \frac16\lf \lambda(N_{00}+\lambda N_{01}),(M_a+ \lambda N_{a0}+\lambda^2 N_{a1}) (M_b + \lambda N_{b0}+\lambda^2N_{b1})\rf_3 v^{a}v^{b}, \notag \\
&& b_4 = \frac{1}{24}\lf \lambda^2
(N_{00}+\lambda N_{01})^{2}, (M_a+ \lambda N_{a0}+\lambda^2 N_{a1}) (M_b + \lambda N_{b0}+\lambda^2N_{b1})\rf_6 v^{a}v^{b}, \notag \\
&& b_5 = \frac{1}{120}\lf 
\lambda^3 (N_{00}+\lambda N_{01})^{3}, (M_a+ \lambda N_{a0}+\lambda^2 N_{a1}) (M_b + \lambda N_{b0}+\lambda^2N_{b1})
\rf_{10} v^{a}v^{b}, \notag \\
&& b_6 = \frac{1}{720}\lf \lambda^4
(N_{00}+\lambda N_{01})^{4}, (M_a+ \lambda N_{a0}+\lambda^2 N_{a1}) (M_b + \lambda N_{b0}+\lambda^2N_{b1})\rf_{15} v^{a}v^{b}, 
\end{eqnarray}

and:
\begin{eqnarray}
&& c_1 = \lf \lambda(N_{ab0}+\lambda N_{ab1})\rf_1 v^{a}v^{b}, \quad
c_2 = \frac{1}{2}\lf \lambda^2(N_{00}+\lambda N_{01}), (N_{ab0}+\lambda N_{ab1})\rf_2 v^{a}v^{b}, \notag \\
&& c_3 = \frac{1}{6}\lf \lambda^3(N_{00}+\lambda N_{01})^{2},(N_{ab0}+\lambda N_{ab1})\rf_3 v^{a}v^{b}, \notag \\
&& c_4 = \frac{1}{24}\lf \lambda^4(N_{00}+\lambda N_{01})^{3},(N_{ab0}+\lambda N_{ab1})\rf_4 v^{a}v^{b}.
\end{eqnarray}

 Separating the terms of order $\lambda^4$, we obtain:
\barr
&& \hspace{-1cm} \frac{ \, b_{2 \, ab}}{\lambda^4} =  \frac12 \lf N_{a1}N_{b1}\rf_1 \, ,
\label{BAB2} \\
&& \hspace{-1cm}\frac{b_{\, 3 \,ab}}{\lambda^4} =  \frac16\Big(\lf N_{00},N_{a0},N_{b1} \rf_{6}+\lf N_{01},M_a,N_{b1} \rf_{6}+\lf N_{01},N_{a0}N_{b0} \rf_{3}\Big), 
\label{BAB3} \\
&& \hspace{-1cm} \frac{b_{\, 4\, ab}}{\lambda^4} =  \frac{1}{24}\Big(\lf(N_{00})^2,M_a,N_{b1}\rf_{12}+\lf(N_{00})^2,N_{a0}N_{b0}\rf_{6} \Big)  \nonumber\\
&&+\lf N_{00},N_{01},M_a,N_{b0}\rf_{24} +\lf (N_{01})^2,M_aM_b\rf_{6}\Big),\label{BAB4} \\
&& \hspace{-1cm}
\frac{b_{\, 5\, ab}}{\lambda^4} = \frac{1}{120}\Big(\lf(N_{00})^3,M_{a},N_{b0}\rf_{20}+\lf (N_{00})^2,N_{01},M_{a}M_{b}\rf_{30}\Big), 
\label{BAB5} \\
&& \hspace{-1cm} \frac{b_{\, 6\, ab}}{\lambda^4} = 
+\frac{1}{720}\lf (N_{00})^4,M_aM_b\rf_{15}\, ,
\label{BAB6}
\earr

and:
\barr
&&
\frac{c_{\, 2 \,  ab}}{\lambda^4} = \frac12\lf N_{01},N_{ab1}\rf_2 \, ,
\label{CAB2} \\
&&\frac{c_{\, 3\, ab}}{\lambda^4} = \frac16\Big(\lf (N_{00})^2,N_{ab1}\rf_3+\lf N_{00},N_{01},N_{ab0}\rf_6\Big), 
\label{CAB3} \\
&&
\frac{c_{\, 4\, ab}}{\lambda^4} = \frac{1}{24}\lf (N_{00})^3,N_{ab0}\rf_4 \, ,
\label{CAB}
\earr
where the contraction with $v^{a}v^{b}$ has been factored out but~taken into account when building new~permutations. 

We now compute the traces in $\gamma^5(b_{ab})$ and $\gamma^5(c_{ab})$   using the fact that every matrix operator  commutes with $\gamma^5$. For~the terms involving $\gamma^5(b_{ab})$, we find: 
\barr
 &&{\rm tr \,} \gamma^5 \lf N_{a1}N_{b1}\rf_1 = 0, \quad  {\rm tr \,} \gamma^5\lf N_{00},N_{a0},N_{b1}\rf_6 = 0,  \\
 &&{\rm tr \,} \gamma^5\lf (N_{00})^2,M_{a},N_{b1}\rf_{12} = 0,   \\
&& {\rm tr \,}\gamma^5\lf (N_{00})^2,N_{a0}N_{b0}\rf_{6} = 0,  
\qquad {\rm tr \,}\gamma^5\lf (N_{00})^3,M_{a},N_{b0}\rf_{20} = 0,
\\
&& {\rm tr \,}\gamma^5\lf (N_{00})^2,N_{01},M_{a}M_{b}\rf_{30} = 0,\qquad {\rm tr \,} \gamma^5\lf (N_{00})^4,M_{a}M_{b}\rf_{15} = 0.
\earr
The only non-zero contributions are: 
\barr
\label{B_1ab}
&&\hspace{-1.5cm} {\rm tr}\, \gamma _{5}\left\lfloor N_{01},M_{a},N_{b1}\right\rfloor
_{6}= k_a k_i \Big(24e^{2}F_{\alpha \beta }F_{4b}\epsilon ^{\alpha \beta
i4}+8ie \epsilon ^{\alpha \beta ij}\left[ F_{\alpha \beta
},D_{b}D_{j}\right] \Big) \equiv {\mathfrak B}_{\, 1\, ab}, \label{100N} \\
&& \hspace{-1.5cm} {\rm tr} \, \gamma _{5}\left[ N_{01},\;N_{a0}N_{b0}\right]
_{3}=-8iek_{a}k_{b}(F_{\mu \nu }D_{i}D_{j}+D_{i}D_{j}F_{\mu \nu
}-D_{i}F_{\mu \nu }D_{j})\epsilon ^{\mu \nu ij} \nonumber 
 \\
&& \hspace{1.5cm}-8iek_{a}k_{j}(\left\{ F_{\mu \nu },\left[ D_{i},D_{b}\right] \right\}
+D_{b}F_{\mu \nu }D_{i}-D_{i}F_{\mu \nu }D_{b})\;\epsilon ^{\mu \nu ij} \equiv {\mathfrak B}_{\, 2\, ab}, \label{101N}
\\
&& \hspace{-1.5cm} {\rm tr} \, \gamma _{5}\left[ \left( N_{01}\right) ^{2}, M_{a}M_{b}\right]
_{6}=16k_{a}k_{b}e^{2}\mathbf{k}^{2}\epsilon ^{\alpha \beta \mu \nu
}F_{\alpha \beta }F_{\mu \nu } \nonumber \\
 &&  \hspace{+1.6cm} +16e^{2}k_{a}k_{b}k_{i}k_{j}\left(4\epsilon ^{\alpha \beta j4}F_{\alpha
\beta }F_{i4}-2F_{\alpha \beta }F_{i\nu }\epsilon ^{\alpha \beta j\nu
}\right) \equiv {\mathfrak B}_{\, 3\, ab},  \label{102N}
\\
&&\hspace{-1.5cm} {\rm tr} \, \gamma _{5}\left[ N_{00}, N_{01}, M_{a}, N_{b0}\right] _{24} = 32ei(k_{a}k_{i}k_{b}k^{\mu })\epsilon ^{\alpha \beta ki}\left[D_{\mu
}D_{k}F_{\alpha \beta }-F_{\alpha \beta }D_{k}D_{\mu }\right]  \nonumber \\
 && \hspace{1.7cm}  + 32ei(k_{a}k_{i}k_{b}k^{\mu })\epsilon ^{\alpha \beta ki}\left[D_{k}F_{\alpha \beta
}D_{\mu }-D_{\mu }F_{\alpha \beta }D_{k}\right] \equiv {\mathfrak B}_{\, 4\, ab}. \label{103N}
\earr

Here, we introduce the notation ${\mathfrak B}_{\,1 \, ab}, {\mathfrak B}_{\, 2\, ab} $, which contribute to the term $b_{\, 3 \, ab}$ together with ${\mathfrak B}_{\,3 \, ab}, {\mathfrak B}_{\, 4\, ab} $, which contribute to the term $b_{\, 4 \, ab}$. 

For the terms involving $c_{ab}$, we have:
\begin{equation}
    {\rm tr} \, \gamma^5\lf N_{01},N_{ab1}\rf_2 = 0, \quad 
    {\rm tr} \, \gamma^5\lf (N_{00})^2,N_{ab1}\rf_3 = 0,
\end{equation}
\begin{equation}
    {\rm tr} \, \gamma^5\lf N_{00},N_{01},N_{ab0}\rf_6 = 0, \qquad 
 {\rm tr}\, \gamma^5\lf (N_{00})^3,N_{ab0}\rf_4 = 0,
\end{equation}
and thus:
\begin{equation}
    {\rm tr} \, \gamma^5\frac{c_{ab}}{\lambda^4} = 0.
\end{equation}

In other words, at~this stage, we have:
\beq
\left(\frac{1}{\lambda^4} {\rm tr} \,  {\gamma^5} e^{\mat{A}+\lambda \mat{B}}\right)_{ab}= \frac{1}{3!} ({\mathfrak B}_{\,1 \, ab}+{\mathfrak B}_{\,2 \, ab})+ \frac{1}{4!} ({\mathfrak B}_{\,3 \, ab}+{\mathfrak B}_{\,4 \, ab}).
\label{107N}
\eeq

\subsection{The Integration over the~Momentum}
\label{INT}
Recalling Eq. (\ref{ANOMALY}), the~last step in our calculation is the integration over the momentum in the expression:
\begin{equation}
\alpha^{(2)}(x) = \int \frac{\mathrm{d}^{4}k}{
(2\pi )^{4}}e^{-k_\mu k_\mu}\left(1-(v^ak_{a})^{2}\right)\left(\frac{1}{\lambda ^{4}}
{\rm tr } \, \gamma^{5}e^{\mat{A}+\lambda \mat{B}} \right)\equiv \frac{1}{16 \pi ^2}(\Delta_1+ \Delta_2),
\label{112N}
\end{equation}
which includes the full contribution to the anomaly in quadratic order in $v^a$,  after~using Eqs. (\ref{100N})--(\ref{103N}) together with Eq. (\ref{107N}).
Once again, we perform the integrals given in Eq. (\ref{integrals}). 

To begin with, we deal with the first term of the right-hand side of Eq. (\ref{112N}), which yields the contribution in the term $\Delta_1$. The~momentum integration provides the intermediate results:
\barr
&&\frac{1}{\pi^2} \int \, d^4 k\ e^{-k_\mu k_\mu} \left(\frac{1}{3!}{\mathfrak B}_{\, 1\, ab}\right)= +\frac{2}{3}ie\epsilon ^{\alpha \beta aj}\left[
F_{\alpha \beta },D_{b}D_{j}\right] \;\;+2e^{2}F_{\alpha \beta
}F_{4b}\epsilon ^{\alpha \beta a4}, \\
&&\frac{1}{\pi^2} \int \, d^4 k\ e^{-k_\mu k_\mu} \left(\frac{1}{3!}{\mathfrak B}_{\, 2\, ab}\right) = +\frac{2}{3}ie\delta _{ab}(ieF_{ij}F_{\mu \nu
}+D_{i}F_{\mu \nu }D_{j})\epsilon ^{\mu \nu ij}  \nonumber \\
&&\hspace{2.5cm}+\frac{2}{3}ie\epsilon ^{\mu \nu aj}(\left\{ F_{\mu \nu },\left[
D_{j},D_{b}\right] \right\} +D_{b}F_{\mu \nu }D_{j}-D_{j}F_{\mu \nu }D_{b}), \\
&&\frac{1}{\pi^2} \int \, d^4 k\ e^{-k_\mu k_\mu} \left(\frac{1}{4!}{\mathfrak B}_{\, 3 \, ab}\right)=
e^{2}\delta _{ab}\frac{3}{4}\epsilon ^{\alpha
\beta \mu \nu }F_{\alpha \beta }F_{\mu \nu }\nonumber \\
&& \hspace{2.5cm} +e^{2}\left( 
\frac{4}{3}%
\epsilon ^{\alpha \beta b4}F_{\alpha \beta }F_{a4}-\frac{2}{3}F_{\alpha
\beta }F_{a\nu }\epsilon ^{\alpha \beta b\nu }\right), \\
&&\frac{1}{\pi^2} \int \, d^4 k\ e^{-k_\mu k_\mu} \left(\frac{1}{4!}{\mathfrak B}_{\, 4\, ab}\right)= -\frac{1}{3}ie\delta _{ab}\epsilon ^{\mu \nu
ij}\left[ +ieF_{\mu \nu }F_{ij}+2D_{i}F_{\mu \nu }D_{j}\right]  \nonumber \\
&&\hspace{2.5cm} -\frac{2}{3}ie\epsilon ^{\mu \nu aj}\left[ F_{\mu \nu
}D_{j}D_{b}-D_{b}D_{j}F_{\mu \nu }+D_{b}F_{\mu \nu }D_{j}-D_{j}F_{\mu \nu
}D_{b}\right]. 
\earr

Combining the above equations and after a long but straightforward calculation, the~potentially dangerous terms containing covariant derivatives $D_i$ either cancel or arrange themselves to produce the commutator $[D_i, D_j]=-ie F_{ij}$. We recall that terms proportional to $k^\mu$ in the integration, arising from Eq. (\ref{103N}), do not contribute when $\mu=4$ due to the antisymmetry in that variable. Furthermore, we remark that we can interchange $a  \leftrightarrow b$ since the result must be symmetric after the multiplication by $v^a v^b$. 

From Eq. (\ref{107N}), we obtain:
\barr
&&\hspace{-1cm}\frac{1}{\pi^2} \int \, d^4 k\ e^{-k_\mu k_\mu} \left(\frac{1}{\lambda ^{4}}
{\rm tr}\, \gamma^{5}e^{\mat{A}+\lambda \mat{B}} \right)_{ab}=\;e^{2}\delta _{ab}\;\left[ \frac{3}{4}\epsilon
^{\mu \nu \alpha \beta }F_{\mu \nu }F_{\alpha \beta }-\frac{1}{3}\epsilon
^{\mu \nu ij}F_{\mu \nu }F_{ij}\right] \nonumber  \\
 && \hspace{+2.5cm} -\frac{2}{3}e^{2}F_{\alpha \beta }F_{b\nu }\epsilon ^{\alpha \beta a \nu }+%
\frac{2}{3}e^{2}\epsilon ^{\alpha \beta aj}F_{\alpha \beta }F_{jb}-\frac{2}{3}%
e^{2}\epsilon ^{\alpha \beta a4}F_{\alpha \beta }F_{b4}.
\label{117N}
\earr

Using the relation:
\beq
\epsilon ^{\mu \nu ij}F_{\mu \nu }F_{ij}=\frac{1}{2}\epsilon ^{\mu \nu
\alpha \beta }F_{\mu \nu }F_{\alpha \beta}
\eeq
in the first line of (\ref{117N}), and {separating into $\nu=4$ and 
$\nu=j$ the summation over $\nu$ in the first term of the second line,} 
results in:
\begin{eqnarray}
\Delta_1 &=&\frac{1}{\pi ^{2}}\int d^{4}ke^{-k_{\mu }k_{\mu }} \left(\frac{1}{\lambda ^{4}}
{\rm tr}\, \gamma^{5}e^{\mat{A}+\lambda \mat{B}} \right)_{ab} v^{a}v^{b}  \nonumber \\
&=&e^{2}\delta _{ab}\;v^{a}v^{b}\frac{7}{12%
}\epsilon ^{\mu \nu \alpha \beta }F_{\mu \nu }F_{\alpha \beta } -\frac{4}{3}e^{2} \epsilon ^{\alpha \beta ib}F_{\alpha \beta
}F_{ia}v^{a}v^{b} -\frac{4}{3}e^{2}F_{ij}F_{a4}\epsilon
^{ijb4}v^{a}v^{b}. 
\end{eqnarray}%

The additional identity:
\beq
v^{a}v^{b}F_{\alpha \beta }F_{ia}\epsilon ^{\alpha \beta
ib}=v^{a}v^{b}F_{ij}F_{4a}\epsilon ^{ijb4}+\frac{1}{4}|\mathbf{v}%
|^{2}\epsilon ^{\mu \nu \alpha \beta }F_{\mu \nu }F_{\alpha \beta },\qquad 
|\mathbf{v}|^{2}=\delta _{ab}\;v^{a}v^{b},
\eeq 
yields the further simplification:
\beq
\Delta_1=\frac{1}{\pi ^{2}}\int d^{4}ke^{-k_{\mu }k_{\mu }} \left(\frac{1}{\lambda ^{4}}
{\rm tr}\, \gamma^{5}e^{\mat{A}+\lambda \mat{B}} \right)_{ab} v^{a}v^{b} =\;\frac{1}{4}e^{2}|\mathbf{v}|^{2}\epsilon
^{\mu \nu \alpha \beta }F_{\mu \nu }F_{\alpha \beta }.
\label{121}
\eeq

The remaining contribution $\Delta_2$ in the expression (\ref{112N}) comes from the zeroth order expansion in $v^a$ of the exponential, which is:

\beq
{\rm tr}\, \gamma _{5}\left( e^{A+\lambda B}\right)=\frac{1}{2!}{\rm tr}\, \gamma^5(\lambda^2 N_{01})^2  =\lambda ^{4}\, \frac{e^{2}}{2}%
\epsilon ^{\mu \nu \alpha \beta }F_{\mu \nu }F_{\alpha \beta},
\eeq
giving:
\barr
\Delta_2 &=&-v^{a}v^{b}\frac{1}{\lambda ^{4}\pi ^{2}}\int d^{4}ke^{-k_{\mu
}k_{\mu }}(k_{b}k_{a})\frac{e^{2}}{2}\epsilon ^{\mu \nu \alpha \beta }F_{\mu
\nu }F_{\alpha \beta }=-|\mathbf{v}|^{2}\frac{e^{2}}{4}\epsilon ^{\mu \nu \alpha \beta
}F_{\mu \nu }F_{\alpha \beta }.
\earr

Summarizing, the~second order contribution in $v^a$ to $\alpha(x)=(\Delta_1 +\Delta_2)/(16\pi^2)$ is identically~zero.

\section{Discussion and~Conclusions}
\label{DISC}

We calculated the abelian axial anomaly in a Lorentz violating model for a  particular case  of the  modification $\gamma^\mu \rightarrow \gamma^\mu + d^\mu{}_\nu \gamma^\nu \gamma^5$  introduced  in the fermionic sector of the SME~\cite{SM2,SM5}. Motivated by the continuum  Hamiltonian of a tilted WSM, whose cones have opposite tilting, we considered the LIV parameter  $d^\mu{}_\nu =\delta^\mu_i \delta ^0_\nu v^i$, where $v^i$ is the tilting parameter given by the microscopic structure of the material. Using the Fujikawa path integral approach~\cite{FUJIKAWA}, we identify the modified axial current $J^\mu_5$ and calculate the anomaly to second order in the tilting parameter. For~a linear order in $v^i$, we find the  rather unexpected gauge invariant result $\alpha^{(1)}(x) \sim ev^j\partial_j \partial_i F^{i0}$, saying that the chiral current obtains the additional term $J^{(1) \mu}_{5}= ev^j\partial_j F^{\mu 0}$. Its contribution to the chiral charge ${Q^{(1)}_5}=\int d^3 x J^{(1)0}_{5} $  is identically zero, which displays this term as an irrelevant contribution to the anomaly. Thus, we conclude that the first order correction to the anomaly is zero. For~the second order correction, we also found a zero contribution, after~a highly non-trivial cancellation of a variety of combinations of spacetime operators which live in matrix space. 
Then, to~the order considered, our result in the Euclidean space is the standard one:
\begin{equation}
    \alpha(x) = \frac{e^2}{32\pi^2}F_{\mu\nu}F_{\rho\sigma}\varepsilon^{\mu\nu\rho\sigma},
\end{equation}
yielding the well-known  anomaly in Minkowski space: 
\beq
\partial_\mu J^\mu_5=- \frac{e^2}{16\pi^2} \, F_{\mu\nu}F_{\rho\sigma}\varepsilon^{\mu\nu\rho\sigma}= \frac{e^2}{2 \pi^2}   \, {\mathbf E} \cdot {\mathbf B}. 
\label{122N}
\eeq

We then comment on  previous work reported in the literature dealing with the calculation of axial anomalies in the Lorentz invariance violating (LIV) case, emphasizing those which use the Fujikawa approach and consider the $d^\mu{}_\nu$ contribution in Eq. (\ref{GAMMAMU}). First, we compare our calculation  with that of Ref.~\cite{KZ},  appropriately expanded to the second order, and~which served as a motivation for the present endeavor. Even though we deal with the same system, and~contrary to our case, the~authors in this reference find non-zero corrections to the standard axial anomaly using the Fujikawa~approach.  

As one can see from our calculation, after~discarding
the irrelevant term proportional to $v^j\partial_j \partial_i F^{i4}$, the~cancellation of the linear and the quadratic contributions heavily rests upon the  appearance of terms including the covariant derivatives $D_\mu$, which must come in the right combinations to finally produce a gauge invariant result, in~accordance with the regularization employed. 
{In our case, the terms $D_\mu D^\mu$ cancel, and  those proportional to $D_i D_j F_{kl},\, D_i F_{kl} D_j,  F_{kl} D_i D_j $, for~example,  manage to yield gauge invariant contributions which are functions of the  electromagnetic tensor.}
We take these facts  as a strong support to the correctness of our evaluation. Such factors involving the covariant derivatives do not appear in Ref.~\cite{KZ}, and we feel  that this is  due to an incorrect separation of the non-commuting terms in the exponential. In~fact, the~non-commuting term ${\mat A}=-2i \gamma^i  \gamma^4 \gamma^5 v_4^jk_j k_i$ cannot be factored out as {$e^{\mat A} \, e^{\lambda {\mat B}}$} in the full expression $e^{{\mat A}+\lambda {\mat B}}$ {, which } is the starting point of \mbox{the~calculation.}

Another representative work is that of Ref.~\cite{FIDEL}, which also considers  the
Fujikawa approach to calculate the LIV abelian axial anomaly  for   a particular case of the contribution $\Gamma^\mu$ in the SME. They  impose  the restriction $d_{\;\nu }^{\mu
}=Q\;c_{\;\nu }^{\mu }$ with the idea of keeping the modified Dirac algebra
as close as possible to the  Lorentz covariant  case. Here, $Q$ is a constant number.
 To  apply their calculation to our case requires expressing our generalized gamma matrix $\Gamma^\mu$ in terms of theirs, which amounts to solving the equation:
\beq
\Gamma^\mu=\gamma^\mu+ \delta^\mu_i v^i \gamma^0 \gamma^5=
(\delta^\mu_\nu+ c^\mu{}_\nu)\gamma^\nu(1+Q\gamma^5).
\eeq

It is a direct calculation to show that:
\beq
c^\mu{}_\nu=0,  \qquad Q=\frac{1}{4} v^i \gamma_i \gamma^0, 
\eeq
thus yielding a matrix-valued $Q$ which lends inapplicable their method to our problem. Nevertheless, the~
 conclusion in Ref.~\cite{FIDEL} is that the corresponding  abelian axial anomaly is not sensitive to the  terms that violate Lorentz  and  that it  is given by the standard Lorentz invariant expression (\ref{122N}). The~authors in  Ref. \cite
{BAETA} also employ the Fujikawa method for the  calculation of the anomaly in the Lorentz violating case,  but~they set $\;d_{\;\nu }^{\mu }=0\;$  from the~outset.

Turning to  the perturbative approach, the~work in Ref.~\cite{SALVIO}
generalizes the standard triangle calculation to include the tensors \ $%
c_{\;\nu }^{\mu }$ and $d_{\;\nu }^{\mu }$\ without any restriction.  The~author deals with the general case of non-abelian chiral theories. His conclusion is that the left and right  chiral anomalies
are independent of the LIV parameters $%
c_{\;\nu }^{\mu }$ and $d_{\;\nu }^{\mu }$, thus keeping the  original form  corresponding to the Lorentz covariant case. In~particular, the~abelian chiral anomaly would be  still given  by Eq. (\ref{122N}). 

 The clash between Ref.~\cite{KZ} and the general result~\cite{SALVIO}, together with the fact 
 that the alternative  methods employed (Fujikawa approach versus perturbative calculation) are not known to be equivalent in the full LIV case, was our main motivation to perform this independent calculation. Our conclusion is also that, to~the order considered, the~calculated abelian axial   anomaly is insensitive to the LIV~corrections.  
 
It is important to emphasize that even though the anomaly turned out to be  independent of the LIV modifications of the fermionic Hamiltonian describing a specific material, this cannot imply  that the 
 electromagnetic response of the  material will also be independent on them. For~example, the~different electromagnetic response of untilted ($d_{\;\nu }^{\mu }=0 $) versus tilted ($d_{\;\nu }^{\mu }\neq 0 $) Weyl semimetals has been experimentally shown in Refs.~\cite{CLZHANG, YYLV}.
 
In fact, the calculation of the effective action using a chiral rotation involving the corresponding   Jacobian, which is related to the anomaly, 
is rather subtle---as clearly shown in Ref.~\cite{ZYUZIN} for the case of a WSM.  We find it illuminating to briefly describe  the main steps of the procedure. The~starting point  
is the Lagragian density:
\beq
{\cal L}=i\bar{\Psi}\left( \gamma ^{\mu }(\partial _{\mu} -ieA_\mu)+ib_{\mu }\gamma ^{\mu
}\gamma ^{5}\right) \Psi,
\eeq
where $b_\mu$ is a four-vector which describes the separation of the Weyl nodes in energy--momentum space. Then, an infinitesimal chiral rotation $\delta \theta (x)=i \, \delta s \,  \gamma_5 \theta(x), \, 0\leq s \leq 1$, with~$\theta(x)=b_\mu x^\mu$ is implemented, which leads to: 
\beq
Z[A_\mu]=\int {\cal D} {\bar \Psi} \,  {\cal D }\Psi \,  \exp  \, \Bigg[\ln J^A+ \int d^4x \,{\bar \Psi} \,  i\gamma ^{\mu }(\partial _{\mu} -ieA_\mu + b_{\mu }(1-\delta s)\gamma _{5}) \Psi  \Bigg].
\eeq

Observe that this rotation introduces both the Jacobian $J^A$ together with the axial current in the term proportional to  $\delta s$. Then, the Jacobian is calculated, yielding: 
\barr
\ln J^A= - \delta s \int d^4 x \,  \theta(x) \, \frac{e^2}{16 \pi^2}   \epsilon^{\mu \nu \alpha \beta} \, F_{\mu\nu} F_{\alpha \beta }.
\label{127N}
\earr

{The final step is the integration of $\delta s$
from zero to one, which eliminates the 
term $b_\mu(1-\delta s) \gamma_5$, and introduces the anomaly contribution (\ref{127N}).}
 This says we can trade the original microscopic information  contained in $b_\mu \gamma^\mu \gamma_5$ by the additional macroscopic effective electromagnetic action originating from $\ln J^A$, thus defining the effective action. In~other words, the~contribution $\ln J^A$ from the Jacobian of the chiral rotation by itself is not necessarily the effective action, unless~we are able to perform steps similar to those shown~above.

A pending goal in this research would be  to obtain the effective electromagnetic action in our approximation and also  as   a non-perturbative result in the tilting parameter $v^i$. This would be especially important in the context of Weyl semimetals, since  the value of $v^i$ will not necessarily as small as it must be in the case of LIV in the SME. Furthermore, the~case $|\bld{v}|=1$ is of special physical importance, as~it corresponds to cones tilted parallel to the Fermi energy plane. This value is also relevant since it distinguishes the so-called type-I ($|\mathbf{v}| < 1$) from type-II Weyl semimetals ($|\mathbf{v}| > 1 $), and it is the point at which the density of states diverges, requiring an additional   regularization~\cite{Soyulanov}. Only a non-perturbative approach in $v^i$ could probe strongly tilted WSMs. A~more complete calculation of the effective action would also necessarily involve the incorporation of the separation of the nodes given by  $b_\mu$ in momentum~space.
\vspace{6pt}

{\bf Author Contributions}: All authors  equally contributed to this work. All authors have read and agreed to the published version of the~manuscript. 

{\bf Funding:} This research was funded by the project DGAPA-UNAM IN103319. Furthermore, we aknowledge support from  CONACyT (M\'exico) under the project Fordecyt-Pronaces- 428214/2020.

{\bf Acknowledgments:} We thank A. Salvio for informative~correspondence. 

{\bf Conflicts of interest:} The authors declare no conflict of interest.

\appendix
\section{Useful Relations}
\label{USEREL}
Here, we collect some identities that we used in the course of the calculation. For~the traces of the Euclidean Dirac matrices, we have:
\begin{equation}
\begin{split}
    & {\rm tr \,} \gamma^5 ={\rm tr \,} \gamma^5 \gamma^{\mu}\gamma^{\nu} = {\rm tr \,} \gamma^5\gamma^{\mu}\gamma^{\nu}\gamma^{\rho} = 0, \\
    & {\rm tr \,} \gamma^{\mu_1}\gamma^{\mu_2}\, .\, .\, .\gamma^{\mu_{2n+1}} = 0, \qquad \mu_n = 1,2,3,4, \\
    & {\rm tr \,} \gamma^{\mu}\gamma^{\nu} = 4\eta^{\mu \nu}, \\
    & {\rm tr \,} \gamma^{\mu}\gamma^{\nu}\gamma^{\rho}\gamma^{\sigma} = 4\big(\eta^{\mu \nu}\eta^{\rho \sigma}+\eta^{\mu \sigma}\eta^{\nu \rho}-\eta^{\mu \rho}\eta^{\nu \sigma}\big), \\
    & {\rm tr \,} \gamma^5\gamma^{\mu}\gamma^{\nu}\gamma^{\rho}\gamma^{\sigma} = -4\varepsilon^{\mu \nu \rho \sigma}, \\
    & {\rm tr \,} \gamma^5\gamma^{\mu}\gamma^{\nu}\gamma^{\rho}\gamma^{\sigma}\gamma^{\alpha}\gamma^{\beta} = -4\big(\eta^{\mu\nu}\varepsilon^{\rho\sigma\alpha\beta}-\eta^{\mu\rho}\varepsilon^{\nu\sigma\alpha\beta}+\eta^{\nu\rho}\varepsilon^{\mu\sigma\alpha\beta} \\
    &\hspace{3.5cm} +\eta^{\alpha\beta}\varepsilon^{\mu\nu\rho\sigma}+\eta^{\sigma\alpha}\varepsilon^{\mu\nu\rho\beta}   -\eta^{\sigma\beta}\varepsilon^{\mu\nu\rho\alpha}\big). 
\end{split}
\end{equation}

For example, the~last relation  gives:
\begin{equation}
\begin{split}
    {\rm tr \,} \gamma^5\gamma^{\rho}\gamma^4\gamma^{\mu}\gamma^{\nu}\gamma^j\gamma^4F_{\mu \nu} & = -4\big(\eta^{\rho4}F_{\mu \nu}\varepsilon^{\mu\nu j4}+2F_{4i}\varepsilon^{\rho ji4}\big) , \\
\end{split}
\end{equation}
which appears when computing the trace of Eq. (\ref{B_1ab}). 

Furthermore, some relations involving the electromagnetic field tensor are:
\barr
&&   F_{\mu\nu}F_{i\sigma}\varepsilon^{\mu\nu i\sigma} = \frac34F_{\mu\nu}F_{\rho\sigma}\varepsilon^{\mu\nu\rho\sigma}, \quad F_{\mu\nu}F_{ij}\varepsilon^{\mu\nu ij} = \frac12F_{\mu\nu}F_{\rho\sigma}\varepsilon^{\mu\nu \rho\sigma}, \\
&& F_{\mu\nu} F_{i4} \epsilon^{\mu\nu i 4}=\frac{1}{4} F_{\mu\nu}F_{\rho\sigma}\varepsilon^{\mu\nu\rho\sigma},   \quad k_ik^{\rho}\varepsilon^{\mu\nu ij}F_{j\rho} = k_jk_i\varepsilon^{\mu\nu jr}F_{ir},\label{B1}\\ 
 &&  F_{\mu\nu}v^{a}v^{b}F_{ia}\varepsilon^{\mu\nu ib} = F_{ij}F_{4a}v^{a}v^{b}\varepsilon^{ijb4}+\frac14|\bld{v}|^2F_{\mu\nu}F_{\rho\sigma}\varepsilon^{\mu\nu\rho\sigma},  
    \label{B2}
\earr
where  the $\rho = 4$ component in the last expression of Eq. (\ref{B1}) does not contribute as it will integrate to~zero.


\end{document}